\documentstyle[prl,aps,psfig]{revtex}

\begin{document}
\twocolumn[\hsize\textwidth\columnwidth\hsize\csname
@twocolumnfalse\endcsname
\title{Criticality in Random Threshold Networks: Annealed Approximation and Beyond}
\author{Thimo Rohlf and Stefan Bornholdt}
\address{Institut f\"ur Theoretische Physik, 
Universit\"at Kiel, Leibnizstrasse 15, D-24098 Kiel, Germany \\
{\rm (Received 7 January 2002)} 
}
\maketitle
\begin{abstract}
Random Threshold Networks with sparse, asymmetric connections show complex dynamical behavior similar to Random Boolean Networks, with a transition from ordered to chaotic dynamics at a critical average connectivity $K_c$. In this type of model - contrary to Boolean Networks - propagation of local perturbations (damage) depends on the in-degree of the sites. $K_c$ is determined analytically, using an annealed approximation, and the results are confirmed by numerical simulations. It is shown that the statistical distributions of damage spreading near the percolation transition obey power-laws, and dynamical correlations between active network clusters become maximal. We investigate the effect of local damage suppression at highly connected nodes for networks with scale-free in-degree distributions. Possible relations of our findings to properties of real-world networks, like robustness and non-trivial degree-distributions, are discussed.

\end{abstract} 
\medskip  
]

\section{Introduction}
Recently, research on complex dynamical networks has become increasingly popular in the statistical physicist's community \cite{Strogatz01,DorogovtsevEvoNet01,AlbertBarabasiReview01}. The variety of fields where
networks of many interacting units are found, e.g. gene regulation, neural systems, food webs, species relationships in biological evolution, economic interactions and the organization of the internet, gives rise to the question of common, underlying dynamical and structural principles, as well as to the study of simple model systems suitable as a theoretical framework for a wide class of complex systems. Such a theoretical approach is provided, for example, by Random Boolean Networks (RBN), originally introduced by Kauffman to model the dynamics of genetic regulatory networks in biological organisms \cite{Kauffman69,Kauffman93}. In this class of models the state $\sigma_i \in \{0,1\}$ of a network node $i$ is a logical function of the states of $k_i$ other nodes chosen at random. The logical functions are also chosen at random, either with equal probability or with a suitably defined bias. A phase transition with respect to the average number $\bar{K}$ of inputs per node is observed at a critical connectivity $K_c = 2$ \footnote{This is the critical value for unbiased Boolean functions; a bias $p$ in the choice of the Boolean functions, where $p$ denotes the mean fraction of 1's in the sites outputs, shifts the critical connectivity to $K_c = 1/[2p(1-p)]$.}: for $\bar{K} < K_c$ a ``frozen'' phase is found with short limit cycles and isolated islands of activity, whereas for $\bar{K} \ge K_c$ one finds a ``chaotic'' phase with limit cycles diverging exponentially with system size, and small perturbations (``damage'') propagating through the whole system. While a more detailed understanding of this transition by means of percolation theory \cite{AlbertBaraBoolper00}  is still in its infancy, theoretical insight primarily was gained by application of the so-called \emph{annealed approximation} introduced by Derrida and Pomeau \cite{DerridaP86} and successive extensions \cite{Bastola96,SoleLuque95,LuqueSole96}; the analytical study of damage spreading by means of this approximation allows for the exact determination of critical points in the limit of large system sizes $N$.\\
Closely related to RBN are Random Threshold Networks (RTN), first studied as diluted, non-symmetric spin glasses \cite{Derrida87} and diluted, asymmetric neural networks \cite{DerridaGZ87,KreeZ87}. Networks of this kind also show complex dynamical behavior similar to RBN at a critical connectivity $K_c$. For the case where the number of inputs per node $K$ is constant, theoretical insight could be obtained in the framework of the annealed approximation \cite{Kuerten88a,Kuerten88b}, but the extension of these results to more realistic szenarios, where the number of inputs per node is allowed to vary, fails because of the complexity of the analytical expressions involved.\\
In this paper we undertake an alternative approach which allows us to calculate the critical value $K_c$ for RTN with discrete weights and zero threshold, but with non-constant number of inputs per node. Starting from a combinatorical investigation, we identify an additional degree of complexity in this type of RTN, which is not found in RBN: damage propagation depends on the $in-degree$ of signal-receiving sites. The introduction of the \emph{average probability of damage propagation} allows us to apply the annealed approximation and leads to the surprising result, that this class of random networks has a critical connectitivity $K_c < 2$, contrary to RBN where one always has $K_c \ge 2$. Numerical evidence is presented supporting our analytical results. Finally, we will give a short outlook on the complexity of phenomena in RTN near $K_c$, which are beyond the scope of the annealed approximation, and we will discuss how our findings could relate to properties of real-world networks, such as robustness \cite{BarkaiLeibler97,AlonSBL99,AlbertJB00,BornholSnep00} and non-trivial degree-distributions \cite{BarabasiAlbert99,Strogatz01}.

\section{Random Threshold Networks}
A Random Threshold Network (RTN) consists of $N$ randomly interconnected binary 
sites (spins) with states $\sigma_i=\pm1$.
For each site $i$, its state 
at time $t+1$ is a function of the inputs it receives from other 
spins at time $t$:
\begin{eqnarray} 
\sigma_i(t+1) = \mbox{sgn}\left(f_i(t)\right) 
\end{eqnarray}  
with 
\begin{eqnarray} 
f_i(t) = \sum_{j=1}^N c_{ij}\sigma_j(t) + h.  
\end{eqnarray}
The $N$ network sites are updated \emph{in parallel}.
In the following discussion
the threshold parameter $h$ is set to zero.
The interaction weights $c_{ij}$ take discrete values $c_{ij}=\pm1$, $+1$ or $-1$ with equal probability. If $i$ does not receive signals
from $j$, one has $c_{ij} = 0$.\\
The in-degree $k_i$ of site $i$ thus is defined as the number of weights $c_{ij}$ with $c_{ij} \ne 0$. If $\bar{K}$ denotes the average connectivity of
the network, for large $N$ the statistical distribution of in- and out-degrees follows a Poissonian: 
 \begin{equation} \mbox{Prob}(k_i = k) = \frac{\bar{K}^k}{k!}\,e^{-\bar{K}} \end{equation}
This corresponds to the case where each weight has equal probability $p = \bar{K}/N$ to take a non-zero value.

\section{Damage Spreading in RTN: dependence on the in-degree $k$}

The most convenient way to distinguish the ordered (frozen) and the chaotic phase of a discrete dynamical network is to study the so-called \emph{damage spreading}: for $\bar{K} < K_c$, a small local perturbation, e.g. changing the state of a single site $\sigma_i \rightarrow - \sigma_i$, vanishes, whereas above $K_c$ it percolates through the network. In RBN, the probability $p_s$ that a site $i$ propagates the damage when a single input $j$ is changed ($1 \rightarrow 0$ or $0 \rightarrow 1$) does not depend on the in-degree $k$ of site $i$: if the possible $2^{2^k}$ Boolean functions of the $k$ inputs are chosen with equal probability, one has $p_s = 1/2$; if the Boolean functions are chosen with a bias $p$, where $p$ denotes the mean percentage of 1's in the output of $i$, one has $p_s = 2p(1 - p)$ \cite{DerridaP86}. Hence, in RBN damage spreading only depends on the \emph{out-degree} of the perturbed site. In RTN, the situation turns out to be more complex: here we find that damage spreading strongly depends on the \emph{in-degree} of the sites. In the following, we will use combinatorical considerations to derive the exact distribution $p_s(k)$ for RTN.
 
Consider a site $i$ having $k$ arbitrary input spins, $k \in \{0,1,2,...,N\}$. Let $k_+$ denote the number of spins equal to $+1$, $k_-$ the number of spins equal to $-1$, hence $k = k_+ + k_-$. The state $\sigma_i(t+1)$ of site $i$ at time $t+1$ is given by eqn. (1) and (2). Let us first calculate the probability $p_s(k)$ that a change of the sign of one arbitrary input spin at time $t$ reverses the sign of $i$'s output at $t+1$, i.e. that it leads to $\sigma_i'(t+1) = - \sigma_i(t+1)$.\\
The number of possible configurations of $k$ spins is $2^k$; as in each of these configurations $k$ spins can be flipped (reversed in sign), the total number of possible spin-flips is
\begin{equation} Z_{total} = k\cdot 2^k. \end{equation} Thus $p_s$ is defined as the number of spin reversals leading to $\sigma_i'(t+1) = - \sigma_i(t+1)$, devided by $Z_{total}$.\\

\begin{figure}[htb]
\let\picnaturalsize=N
\def\picsize{85mm}
\def\picfilename{fig1.eps}
\ifx\nopictures Y\else{\ifx\epsfloaded Y\else\input epsf \fi
\let\epsfloaded=Y
\centerline{\ifx\picnaturalsize N\epsfxsize \picsize\fi
\epsfbox{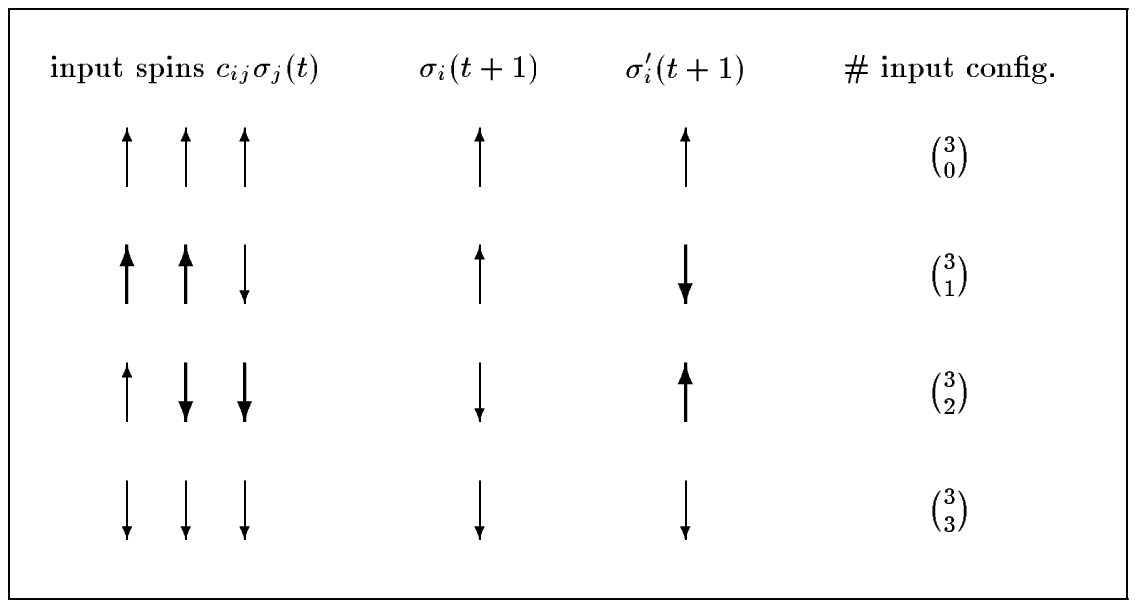}}}\fi
\caption{\label{combideriv}Combinatorical derivation of $p_s(k)$, here for $k=3$. Reversal of the thick-lined inputs $c_{ij}\sigma_j(t)$ leads to a spin-flip of site $i$, i.e. $\sigma_i'(t+1) = - \sigma_i(t+1)$ (thick-lined vectors). The total number of input configurations is 8 (see the right column showing the corresponding numbers of spin configurations), hence the number of possible spin-flips is $3\cdot 8 =24$. The total number of input-spin reversals leading to damage propagation is $2\cdot ( 3 + 3 ) = 12$. Thus we find $p_s(3) = 1/2$. The generalization for  $k = 5,7,9,...$ is straight-forward.}
\end{figure}

\begin{figure}[htb]
\let\picnaturalsize=N
\def\picsize{95mm}
\def\picfilename{fig2.eps}
\ifx\nopictures Y\else{\ifx\epsfloaded Y\else\input epsf \fi
\let\epsfloaded=Y
\centerline{\ifx\picnaturalsize N\epsfxsize \picsize\fi
\epsfbox{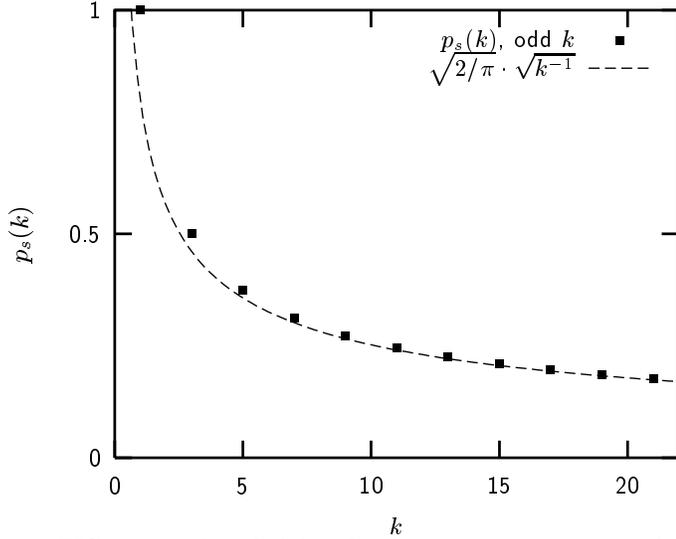}}}\fi
\caption{\label{locflip}Local probability for damage propagation, $p_s(k)$, as a function of the in-degree $k$ of the signal-receiving site, for odd $k$. The squares correspond to the exact result of eqn. (7), the dashed curve shows the Stirling approximation of eqn. (13).}
\end{figure}

For $k=1$ it is easy to see that $p_s=1$, whereas for $k=2$ one gets $p_s=1/2$. For $k\ge 3$ we have to analyze even and odd $k$ seperately:\\\\
\bfseries Odd $k$: \mdseries By the definition of the transition function for site $i$ (eqn. (2)) one recognizes that there are only two configurations in which a flip of a single input spin at time $t$ can lead to a spin-flip of site $i$ at time $t+1$; these configurations are given by the condition:
\begin{equation} |k_+-k_-| = 1. \end{equation}
If $k_+ = k_-+1$ (i.e.  $\sigma_i(t+1)=1$) , the reversal of a positive input spin always leads to $\sigma'_i(t+1)=-1$, whereas flips of negative input spins do not change the state of site $i$ (for $k_+ = k_- -1$ it is vice versa). Thus in both configurations $(k+1)/2$ spin flips of $k$ possible lead to $\sigma_i'(t+1) = - \sigma_i(t+1)$. The total number of configurations fulfilling eqn. 5 is
\begin{eqnarray}  Z_{oddflip} =  { k \choose (k-1)/2} +   { k \choose (k+1)/2}
=  2\cdot{ k \choose (k+1)/2}.\end{eqnarray} 
Hence the number of spin-flips leading to damage spreading at site $i$ is $Z_{oddflip}\cdot(k+1)/2$ and we obtain for $k=3,5,7,...$
\begin{equation} p_s(k) = \frac{Z_{oddflip}\cdot(k+1)}{2\cdot Z_{total}} = \frac{(k+1)\cdot\displaystyle{ k \choose (k+1)/2}}{k\cdot 2^k}\end{equation}
Fig. \ref{combideriv} demonstrates the above derivation at the example $k = 3$. $p_s(k)$ for odd $k$ is shown in Fig. \ref{locflip}.

\bfseries Even $k$: \mdseries For even $k$, there are also only two configurations in which changing a single input of site $i$ at time $t$ leads to $\sigma_i'(t+1) = - \sigma_i(t+1)$, given by the conditions
\begin{equation} k_+ = k_- \quad (*)\quad \mbox{or}\quad k_- = k_+ + 2\quad (**). \end{equation}
In the case $ k_+ = k_- + 2$ neither the reversal of a positive nor the reversal of a negative input-spin changes the output of $i$, due to the signum function in eqn. (2). In the case $(*)$, $k/2$ spin-flips ($+1 \rightarrow -1$) change the output of site $i$ ($+1 \rightarrow -1$), in the case $(**)$ $(k+2)/2$ spin-flips ($-1 \rightarrow +1$) do so. The number of configurations fulfilling eqn. (8) is given by
\begin{eqnarray} Z_{evenflip} =  Z(*) 
+ Z(**) = { k \choose k/2} +   { k \choose (k+2)/2} \end{eqnarray}
So we finally obtain for $k=2,4,6,..$
\begin{eqnarray} p_s(k) &=& \frac{k\cdot Z(*) + (k+2)\cdot Z(**) }{ 2 \cdot Z_{total} }\\
&=& \frac{\displaystyle k\cdot\displaystyle{ k \choose k/2} +\displaystyle (k+2)\cdot\displaystyle{ k \choose (k+2)/2} }{k\cdot 2^{k+1}}\\
&=& p_s(k+1),\end{eqnarray}
as can be seen by some simple algebra. 
Using the Stirling formula for $k!$, one recognizes that the asymptotic behavior of $p_s(k)$ for odd $k$ is given by
\begin{eqnarray} p_s(k) &=& \frac{1}{k+1}\cdot\sqrt{\frac{2\,k}{\pi}}
\cdot\left(\frac{k}{\sqrt{k^2-1}}\right)^k \approx \sqrt{\frac{2}{\pi}}\cdot \frac{1}{\sqrt{k}}. \end{eqnarray}
Thus it turns out that
\begin{equation} \lim_{N \to \infty}\lim_{k \to N} p_s(k) = 0. \end{equation}
However, this does \emph{not} mean that there is no damage spreading for $k \to N$: in each timestep, damage increases $\propto k\cdot p_s(k) \sim \sqrt{k} \sim \sqrt{N}$ in this limit, which ensures the existence of a ``chaotic'' regime.

\section{Average probability for damage spreading}

Choosing an arbitrary network site $i$ with at least one input and changing the sign of one input-spin will lead to a spin-flip of site $i$ with probability $p_s(k)$, dependent on the number of inputs $k$. Now we are interested in the \emph{expectation value} of this probability, i.e. the  average probability for damge spreading $\langle p_s\rangle (\bar{K})$ for a given average network connectivity $\bar{K}$. For large $N$, this problem is equivalent to connecting a new site $j$ with state $\sigma_j$ to an arbitrary site $i$ of a network of size $N-1$ with average connectivity $\bar{K}$; thus $k_i$ is increased by one. Hence, changing the sign of $\sigma_j$ at time $t$ will lead to a different output of site $i$ at time $t+1$ with probability $p_s(k_i+1)$. Using the Poisson approximation and taking the thermodynamic limit $N \to \infty$, this leads to
\begin{equation} \langle p_s\rangle (\bar{K}) =  e^{-\bar{K}}\sum_{k=0}^{\infty} \frac{\bar{K}^k}{k!}\cdot p_s(k+1) \end{equation}
Splitting the sum for even and odd $k$ yields
\begin{eqnarray}
 \langle p_s\rangle (\bar{K}) = e^{-\bar{K}}\left\{ 1 + \sum_{i=1}^{\infty}\frac{\bar{K}^{2i-1}}{(2i-1)!}\cdot p_s(2i) \right. \nonumber\\ 
 \left. +\sum_{i=1}^{\infty}\frac{\bar{K}^{2i}}{(2i)!}\cdot p_s(2i+1)\right\}  \end{eqnarray}
Using the relation $p_s(k) = p_s(k+1)$ for even $k$ one obtains
\begin{eqnarray}
 \langle p_s\rangle (\bar{K}) =  e^{-\bar{K}}\cdot\left\{1 + \sum_{i=1}^{\infty}\bar{K}^{2i-1} p_s(2i+1)\cdot \right. \nonumber\\
\cdot \left. \left(\frac{1}{(2i-1)!} + \frac{\bar{K}}{(2i)!}\right)\right\} \end{eqnarray} 
Inserting (7) into this equation finally yields
\begin{eqnarray}  \langle p_s\rangle (\bar{K}) = e^{-\bar{K}}\left\{ 1 + \frac{1}{2}\sum_{i=1}^{\infty} \frac{1}{(i!)^2}\left(\frac{1}{2i} + \bar{K}\right)\cdot\right. \nonumber \\
\cdot\left. \left(\frac{\bar{K}}{2}\right)^{2i-1}\right\}.
\end{eqnarray}
Formula (18) is the central result of this paper, as it allows for an analytic calculation of the critical connectivity $K_c$ of RTN. For practical use it is worth to notice that the sum in eqn. (18) converges very fast, setting the upper limit to ten is sufficient for a relative error $\Delta \langle p_s\rangle / \langle p_s\rangle < \cal{O}(10^{-4})$.

\begin{figure}[htb]
\let\picnaturalsize=N
\def\picsize{85mm}
\def\picfilename{fig3.eps}
\ifx\nopictures Y\else{\ifx\epsfloaded Y\else\input epsf \fi
\let\epsfloaded=Y
\centerline{\ifx\picnaturalsize N\epsfxsize \picsize\fi
\epsfbox{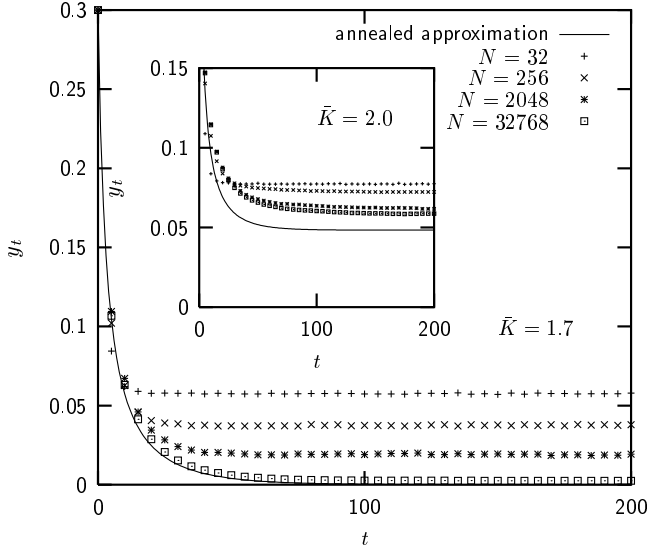}}}\fi
\caption{\label{hadit}Time evolution of the relative Hamming distance $y_t$ of two spin configurations with $y_0 = 0.3$ for different network sizes $N$ with $\bar{K} =1.7$ (numerical simulations, ensemble statistics over 10000 networks / 400 networks ($N=32768$)). For large $N$, one finds convergence against the prediction of the annealed approximation (solid curve). The inset shows the same for overcritical networks ($\bar{K} = 2.0$).} 
\end{figure} 

\section{Calculation of $K_c$ by Derrida's Annealed Approximation}

The still most powerful analytical approach to describe damage spreading in random networks of automata - and thus to calculate the critical value $K_c$ of $\bar{K}$ - is the so-called \emph{annealed approximation} introduced by Derrida and Pomeau \cite{DerridaP86}. This approximation, originally applied to Kauffman networks (Boolean networks with constant number $K$ of inputs per node),  neglects the fact that the (Boolean) functions and the interactions between the spins are quenched, i.e. constant over time, and instead randomly reassigns inputs and functions to all spins at each time step. 
As a central result of this approximation Derrida and Pomeau derived a recursion formula which describes the time evolution of the Hamming distance $d(\vec{\sigma},\vec{\sigma}')$ of two spin configurations $\vec{\sigma}$ and $\vec{\sigma}'$. The normalized distance $y = d(\vec{\sigma},\vec{\sigma}')/N $ at time $t+1$ for $p_{s} = 1/2$ and in-degree $k$ is given by the recursion
\begin{equation} y_{t+1} = \frac{1-(1-y_t)^k}{2}. \end{equation}
They also give a straight-forward generalization to networks where $k$ is not constant: 
\begin{equation} y_{t+1} = \sum_k \rho_k \frac{1-(1-y_t)^k}{2}, \end{equation} 
where $\rho_k$ is the probability that a network site has $k$ inputs.
In the case of an average probability for damage propagation $\langle p_{s}\rangle (\bar{K})$ depending on $\bar{K}$, as for the RTN discussed in this article, this generalizes to 
\begin{equation} y_{t+1} = \langle p_{s}\rangle (\bar{K})\sum_k \rho_k \cdot[1-(1-y_t)^k]. \end{equation}
In the following, we focus our discussion on random networks, where each link has equal probability to take a non-zero value (as first introduced by Erd\"os and R\'{e}nyi \cite{ErdosRenyi59}). Thus, for average connectivities $\bar{K} \ll N$, $\rho_k$ is given by a Poissonian, leading to 
\begin{eqnarray}  y_{t+1} &= & \langle p_{s}\rangle (\bar{K})\, e^{-\bar{K}}\,\sum_k \frac{\bar{K}^k}{k!}\cdot [1-(1-y_t)^k] \nonumber \\
&=& \langle p_{s}\rangle (\bar{K})\, e^{-\bar{K}}\,
\left ( \sum_k \frac{\bar{K}^k}{k!} - \sum_k \frac{\bar{K}^k}{k!}\cdot (1-y_t)^k \right )\nonumber \\
&=&  \langle p_{s}\rangle (\bar{K})\,\left ( 1 - e^{-\bar{K}}\,\sum_k \frac{[\bar{K}\,(1-y_t)]^k}{k!} \right )\nonumber \\
&=&  \langle p_{s}\rangle (\bar{K})\,(1-e^{-\bar{K}y_t}). \end{eqnarray}
Fig. \ref{hadit} compares the time evolution of the Hamming distance, measured in numerical simulations of RTN of different size $N$, to the prediction of eqn. (22).

\begin{figure}[tb]
\let\picnaturalsize=N
\def\picsize{85mm}
\def\picfilename{fig4.eps}
\ifx\nopictures Y\else{\ifx\epsfloaded Y\else\input epsf \fi
\let\epsfloaded=Y
\centerline{\ifx\picnaturalsize N\epsfxsize \picsize\fi
\epsfbox{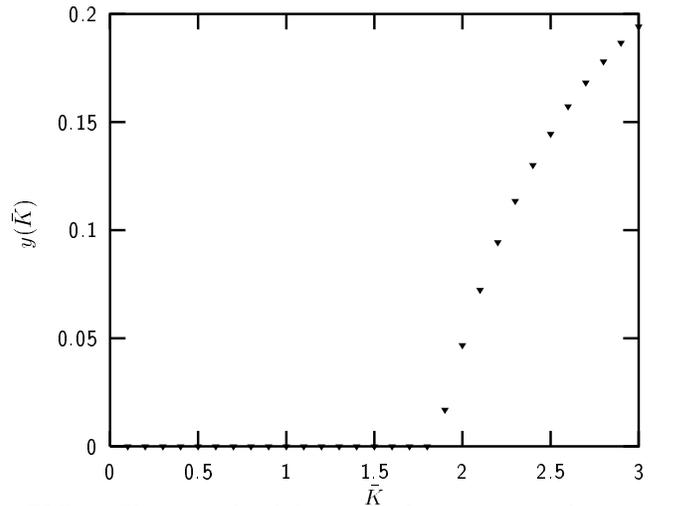}}}\fi
\caption{\label{deri}The normalized distance $y$ for $t \to \infty$ as a function of $\bar{K}$, calculated by numerical solution of eqn. (23). One finds $y(\bar{K}) > 0$ for $\bar{K} \ge K_c =1.849$. }
\end{figure}

\begin{figure}[t]
\let\picnaturalsize=N
\def\picsize{85mm}
\def\picfilename{fig5.eps}
\ifx\nopictures Y\else{\ifx\epsfloaded Y\else\input epsf \fi
\let\epsfloaded=Y
\centerline{\ifx\picnaturalsize N\epsfxsize \picsize\fi
\epsfbox{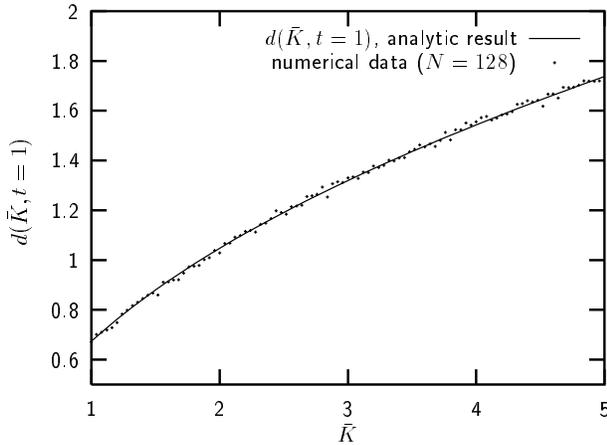}}}\fi
\caption{\label{hadi1}Average Hamming distance $d_t$  for two configurations $\vec{\sigma}$ and $\vec{\sigma}'$ differing in one bit at time $t-1$ as a function of the average connectivity (ensemble average over 10000 random networks with $N=128$ for each data point). One finds excellent agreement with the prediction of the annealed approximation (solid curve, $d_{t, annealed} = \langle p_s\rangle(\bar{K})\cdot \bar{K}$).}
\end{figure}

As well for subcritical as for supercritical $\bar{K}$  one finds convergence against the annealed approximation, but for $\bar{K} \ge  K_c$ the convergence is quite slow (see section 6).\\
In the limit $t \to \infty$ we expect that the normalized distance $y(\bar{K})$ evolves to a constant value, i.e. $y_{t+1} = y_t = y$.
Inserting this condition in eqn. (22) leads to the fixed point equation
\begin{equation} f(y) \equiv y - \langle p_{s}\rangle (\bar{K})\,(1-\exp{[-\bar{K}\cdot y}]) = 0. \end{equation}
Solutions $y^*$ of eqn. (23) for some values of the average connectivity $\bar{K}$ are shown in Fig. 4; evidently, there exists a critical
connectivity $K_c$ above which the Hamming distance of two initially nearby trajectories increases to a non-zero value, indicating damage spreading
through the network. The exact value $K_c$ can be obtained easily by a linear stability analysis of eqn. (23). 
$y^*_0 = 0$ is always a solution of (23), but is attractive (stable) only for $\bar{K} < K_c$. \\
This fixed point is stable only if 
\begin{equation} \lim_{\varepsilon \to 0} \left |\frac{d f(y)}{dy}\right |(y^*_0 + \varepsilon) < 1. \end{equation}
One has
\begin{equation} \left |\frac{d f(y)}{dy}\right |(y^*_0 + \varepsilon) =  \langle p_{s}\rangle (\bar{K})\cdot\bar{K}\cdot\exp{[-\bar{K}\cdot \varepsilon]},\end{equation}
i.e.
\begin{equation} \lim_{\varepsilon \to 0} \left |\frac{d f(y)}{dy}\right |(y^*_0 + \varepsilon) =  \langle p_{s}\rangle (\bar{K})\cdot\bar{K}. \end{equation}
Inserting this result into (24) we find that $y^*_0 = 0$ is attractive if
\begin{equation}  \langle p_{s}\rangle (\bar{K})\cdot\bar{K} < 1.\end{equation}
Thus the critical connectivity $K_c$ can be obtained by solving the equation
\begin{equation}  \langle p_{s}\rangle (K_c)\cdot K_c = 1, \end{equation}
$\langle p_{s}\rangle(\bar{K})$ given by eqn. (18).\\
Eqn. (27) and (28) have a simple interpretation in terms of damage spreading: $\langle p_{s}\rangle (\bar{K})\cdot\bar{K}$ is the expectation value $\langle d_t \rangle$ of the Hamming distance - i.e. of the damage - at time $t$, if at time $t-1$ \emph{a single spin is reversed}, as this spin on average has $\bar{K}$ outputs to other spins, which all will propagate damage with probability $\langle p_{s}\rangle(\bar{K})$. Fig. \ref{hadi1} shows that the prediction $\langle d_{t,annealed} \rangle$ of our annealed approximation agrees perfectly well with with the average $d_t$ measured between two spin configurations in RTN with $d_{t-1} = 1$ (ensemble statistics).\\
A numerical solution of eqn. (28) for the RTN discussed here yields the critical value
\begin{equation} K_c = 1.849 \pm 0.001. \end{equation}
Thus we get the remarkable result that in RTN with $h=0$ and discrete interaction weights, marginal damage spreading - i.e. the percolation transition from frozen to chaotic dynamics - is found \emph{below} $\bar{K} = 2$, in contrast to RBN where one always has $K_c \ge 2$.

\section{Beyond the Annealed Approximation: Complexity in RTN}

\begin{figure}[tb]
\let\picnaturalsize=N
\def\picsize{85mm}
\def\picfilename{fig6.eps}
\ifx\nopictures Y\else{\ifx\epsfloaded Y\else\input epsf \fi
\let\epsfloaded=Y
\centerline{\ifx\picnaturalsize N\epsfxsize \picsize\fi
\epsfbox{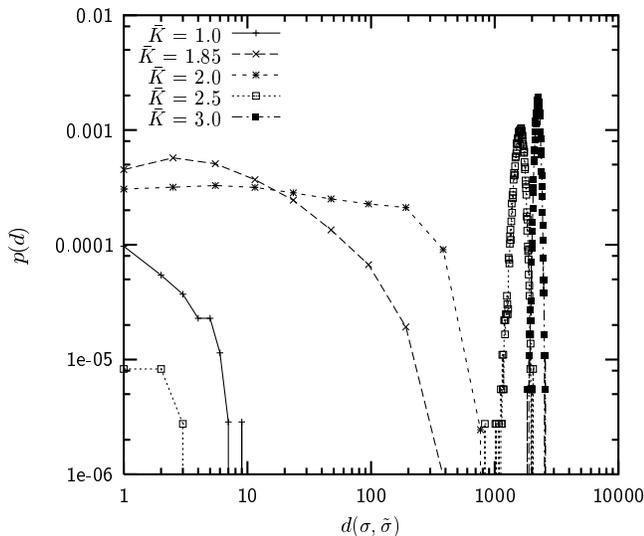}}}\fi
\caption{\label{hadisr}Statistical distributions $p(d)$ of the Hamming distances $d(\sigma, \tilde{\sigma})$ for two spin configurations $\sigma$, $\tilde{\sigma}$ with $d(t=0) = 0.3\cdot N$. Only the tails of the distributions are shown in this log-log-plot. Ensemble statistics was taken after 100 dynamical updates over 5000 RTN with size $N = 8192$ and Poissonian distributed in- and out-degree, testing 20 different initial conditions for each network. In the ordered regime ($\bar{K} = 1$) damage is suppressed exponentially. Near criticality and slightly above ($\bar{K} = 1.85, \bar{K} = 2.0$) the tails aproximately obey power-laws (log-binned data). For $2.1 < \bar{K} < 2.6$, the distributions are bimodal (shown here for $\bar{K} = 2.5$), with an increasing maximium at large damage values. For larger $\bar{K}$, the tails of the distributions become Gaussian, with a maximum approaching $d = N/2$ with increasing $\bar{K}$. Not visible in this plot is the pronounced maximum of all distributions at $d = 0$: even for $\bar{K} = 3.0$ damage finally becomes zero for about $56\%$ of the networks/ the initial conditions tested; this maximum vanishes for $\bar{K} \to N$.}
\end{figure}

So far we presented numerical evidence that the annealed approximation correctly predicts the \emph{average} damage spreading behavior in RTN and thus the critical connectivity $K_c$ in the limit of large system sizes, however, this coarse-grained approach of course does not capture the whole complexity of the network dynamics near criticality. Concerning the statistics of damage spreading, even for quite large $N$ scale-free distributions are found in a certain  range around $K_c$; skewed, super-critical distributions, with an increasing  maximum moving towards $N/2$ with increasing $\bar{K}$, are found for $\bar{K} \ge 2.1$ ($N = 8192$, Fig. \ref{hadisr}). In the ordered regime, the distributions decay exponentially. Presumably, the good convergence of the average Hamming distance measured in numerical simulations for $\bar{K} < K_c$ against the annealed approximation directly reflects averaging over these exponential distributions with well-defined characteristic scale, whereas averaging over the the scale-free damage-distributions around $K_c$ leads to the observed weak finite-size scaling $\propto 1/\log{(N)}$ against the analytical result (Fig. \ref{hadit}) at the order-chaos transition. Furthermore, for finite $N$, even the transition from critical (scale-free) to supercritical distributions (Gaussian distributions in the limit $\bar{K} \to N$) is rather smooth: even deep in the ``chaotic'' regime (e.g. $\bar{K} = 2.5$, Fig. \ref{hadisr}) damage becomes zero for more than $60\%$ of the initial conditions/ the networks tested, and the damage distributions are clearly bimodal.\\
Attractor periods of RTN are found to be power-law distributed in a certain range of connectivities around $K_c$, due to dynamical correlations between network sites, which are neglected completely in the framework of Derrida's annealed approximation. We shall briefly discuss this:
The \emph{average correlation} $\mbox{Corr}(i,j)$ of a pair $(i,j)$ of sites is defined as the average over the products $\sigma_i(t)\sigma_j(t)$ in dynamical network evolution between two distinct points of time $T_1$ and $T_2$:
\begin{equation} \mbox{Corr}(i,j) = \frac{1}{T_2 - T_1}\sum_{t=T_1}^{T_2} \sigma_i(t)\sigma_j(t) \end{equation}  
The \emph{global average correlation} can be defined as
\begin{equation} \langle \mbox{Corr} \rangle_g = \frac{1}{N(N-1)}\sum_{i=1}^{N}\sum_{j=1}^{N}|\mbox{Corr}(i,j)| \quad \mbox{with } i \ne j,\end{equation}
neglecting trivial auto-correlations.

\begin{figure}[tb]
\let\picnaturalsize=N
\def\picsize{85mm}
\def\picfilename{fig8.eps}
\ifx\nopictures Y\else{\ifx\epsfloaded Y\else\input epsf \fi
\let\epsfloaded=Y
\centerline{\ifx\picnaturalsize N\epsfxsize \picsize\fi
\epsfbox{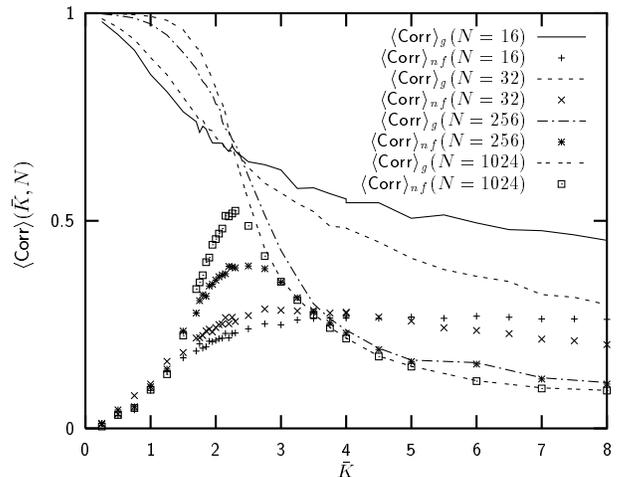}}}\fi
\caption{\label{globcorr}Global average correlation $\langle \mbox{Corr}\rangle_g$ (lines) and average correlation of non-correlated pairs $\langle \mbox{Corr}\rangle_{nc}$ (crosses or points) as a function of the average connectivity $\bar{K}$ for four different system sizes $N$. For each value of $\bar{K}$ sampled ensemble-averages were taken over 1000 RTN. For each network the average was taken over a subset of 5000 randomly chosen pairs of sites. Individual runs were limited to $T_{max} = 20000$.}  
\end{figure}

In the ordered (frozen) regime, there are few, isolated (non-frozen) clusters with dynamical activity, but their dynamics is \emph{not correlated}. A good measure here is the \emph{average correlation of non-correlated sites} $\langle \mbox{Corr} \rangle_{nc}$, i.e. the average correlation of pairs of sites with $0 \le \mbox{Corr}(i,j) < 1$ (Fig. \ref{globcorr}). $\langle \mbox{Corr} \rangle_{nc}$ is almost zero in the ordered regime. Due to this uncorrelated dynamics of a few ``active islands'' in a frozen network the distributions of attractor periods $P$ decay $\propto \exp{(-P)}$. Near $K_c$, however,  $\langle \mbox{Corr} \rangle_{nc}$ shows a pronounced maximum, i.e. the dynamics of active clusters becomes strongly correlated; thus, at this percolation transition, attractor periods show scale-free distributions. In the chaotic regime, damage spreading destroys most of the correlations, consequently, $\langle \mbox{Corr} \rangle_{nc}$ decays once again.

\section{Discussion}

We investigated damage spreading in Random Threshold Networks (RTN) with zero threshold and discrete weights. This kind of discrete dynamical network shows complex dynamics similar to Boolean networks, with a transition from ordered to chaotic dynamics at a critical average connectivity $K_c$. Using combinatorial methods, the exact distribution $p_s(k)$ for local damage propagation was derived. It was shown that Derrida's annealed approximation can be applied to this class of models; this theoretical analysis yielded the surprising result $K_c = 1.849$, in contrast to RBN, where $K_c \ge 2$. The ansatz proposed in this paper could offer a road-map for an analytical treatment of similar systems with additional complexity (non-discrete weights, $h \ne 0$).\\
An interesting result of our studies is that in dynamical networks with sparse asymmetric interactions and some kind of threshold characteristics governing dynamics, a high number $k$ of inputs per node \emph{can stabilize dynamics}, as the effect of single, local errors is reduced like $1/\sqrt{k}$; this, however, is counteracted by the overall topological randomness in the network, which allows for ordered dynamics only at small \emph{average} connectivities $\bar{K}$. This effect of ``local damage suppression'' at nodes with high in-degree ist demonstrated in Fig. 8, directly comparing damage distributions in RTN and RBN: damage spreading was investigated for networks with ``flat'', scale-free in-degree distributions $\propto k_{in}^{-\alpha}$, wheras the out-degree follows a Poissonian. For very flat in-degree distributions ($\alpha = 1$), both RTN and RBN have a hierarchical, clustered structure and show ordered dynamics, but in RTN damage is stronger suppressed than in RBN, due to local damage suppression at highly connected nodes. 

\begin{figure}[t]
\let\picnaturalsize=N
\def\picsize{85mm}
\def\picfilename{fig7.eps}
\ifx\nopictures Y\else{\ifx\epsfloaded Y\else\input epsf \fi
\let\epsfloaded=Y
\centerline{\ifx\picnaturalsize N\epsfxsize \picsize\fi
\epsfbox{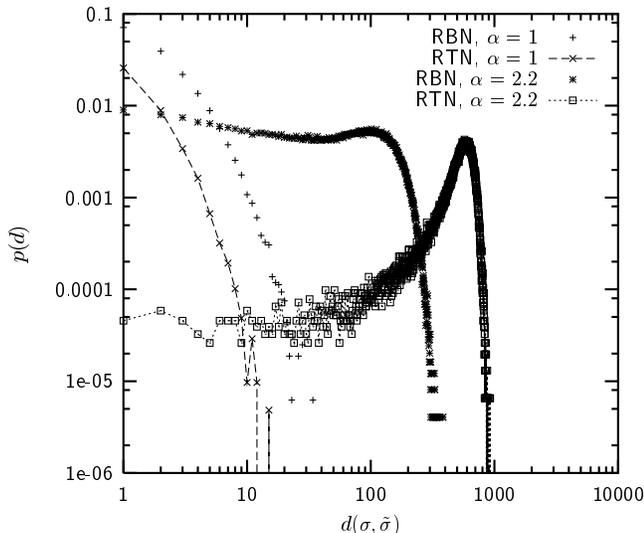}}}\fi
\caption{\label{hadidampow}Statistical distributions $p(d)$ of the Hamming distances $d(\sigma, \tilde{\sigma})$ for two spin configurations $\sigma$, $\tilde{\sigma}$ with $d(t=0) = 0.3\cdot N$. Only the tails of the distributions are shown in this log-log-plot. Ensemble statistics was taken after 100 dynamical updates over 10000 RTN or RBN, respectively, with size $N = 2048$  and average connectivity $\bar{K} = 2.5$, Poissonian distributed out-degree and in-degree-distribution $p(k_{in}) \propto k_{in}^{-\alpha}$. The choice $\alpha = 1$ leads to clustered, hierarchical networks with ordered dynamics, but in RTN damage spreading is stronger suppressed: $p(d)$ shows a faster exponential decay than in RBN. This is a direct effect of the local damage suppression due to highly connected nodes in RTN. For $\alpha = 2.2$, the opposite is observed: Whereas RBN show an only slightly overcritical $p(d)$ with a power-law regime prevailing over almost two decades, RTN show a strongly skewed, overcritical $p(d)$. In this regime the local damage suppression effect $\propto \sqrt{k_{in}}$ is not strong enough to counteract chaotic dynamics increasing $\propto k_{out}$. }
\end{figure}

For steeper in-degree distributions ($\alpha = 2.2$ in Fig. 8) this effect is not strong enough, and the dynamics in RTN is even ``more chaotic'' in RTN than in RBN. Concerning models including \emph{self-organization of network topology} this could have important consequences. Topological evolution of RTN by a local coupling of control- and order parameters (connectivity and magnetization/correlations of network sites) was introduced in \cite{BornholRohlf00}. Recent, more detailed studies of these models show that the self-organization processes, balancing the networks in a regime of complex, non-chaotic dynamics near criticality, indeed favor highly-connected nodes to some extent, leading to deviations of the statistical distribution of in-degrees from a Poissonian \cite{RohlfBornhol01}. Many networks in nature are characterized by non-Poissonian degree distributions (namely power-laws). Whereas for fast-growing networks like the internet models based on preferential linking provide explanations for the observed distributions \cite{BarabasiAlbert99}, for other dynamical networks - e.g. protein networks \cite{JeongMBO01} or gene networks \cite{ThieffryHPC98} - convincing models are still not available; szenarios based on preferential linking have to make detailed a-priori-assumptions about the evolutionary processes leading to the observed structures, which in the case of biological networks usually cannot be falsified. Approaches based on network dynamics \cite{BornholSnep00,BornholRohlf00}, setting the focus on the \emph{robustness} of network-dynamics and -evolution,  could provide more realistic szenarios, as some threshold characteristics usually is an intrinsic dynamical feature of these networks. We expect that future research will underline the significance of RTN as simple ``toy systems'' yet able to capture the essentials of natural dynamical networks: evolution of high robustness and non-trivial randomness.

\section{Acknowledgements} 
T. Rohlf would like to thank the Studienstiftung des deutschen Volkes for financial support
of this work.

\end{document}